\documentclass[aps,prb,twocolumn,longbibliography,groupedaddress,showpacs,floatfix,superscriptaddress,a4paper]{revtex4-2}

\usepackage[dvipsnames]{xcolor}
\definecolor{darkerBlue}{RGB}{30, 81, 149}
\usepackage[
    plainpages=false,
    pdfpagelabels,
    colorlinks=true,
    linkcolor=BrickRed,
    urlcolor=darkerBlue,
    citecolor=darkerBlue,
    pdftitle={Specific heat and density anomaly in the Hubbard model},
    pdfauthor={M. A. Habitzreuter, Willdauany C. de Freitas Silva, Eduardo O. Rizzatti, Thereza Paiva, and Marcia C. Barbosa},
    pdfdisplaydoctitle=true,
    pdfduplex=DuplexFlipLongEdge
]{hyperref}

\usepackage{amsmath,amssymb}
\usepackage{bm}
\usepackage{braket}
\usepackage{orcidlink}
\usepackage{graphicx}

\begin{document}

\title{Specific heat and density anomaly in the Hubbard model}

\author{M. A. Habitzreuter \orcidlink{0000-0002-3534-1409}}
\affiliation{International Institute of Physics, Universidade Federal do Rio Grande do Norte, 59078-970 Natal, Brazil}

\author{Willdauany C. de Freitas Silva \orcidlink{0000-0002-4133-2234}}
\affiliation{Instituto de Física - Universidade Federal do Rio de Janeiro, Rio de Janeiro, RJ 21941-972, Brazil}

\author{Eduardo O. Rizzatti \orcidlink{0000-0003-0845-0105}}
\affiliation{Instituto de Física - Universidade Federal do Rio Grande do Sul, Porto Alegre, RS 91501-970, Brazil}

\author{Thereza Paiva \orcidlink{0000-0002-4199-3809}}
\email{tclp@if.ufrj.br}
\affiliation{Instituto de Física - Universidade Federal do Rio de Janeiro, Rio de Janeiro, RJ 21941-972, Brazil}

\author{Marcia C. Barbosa \orcidlink{0000-0001-5663-6102}}
\email{marcia.barbosa@ufrgs.br}
\affiliation{Instituto de Física - Universidade Federal do Rio Grande do Sul, Porto Alegre, RS 91501-970, Brazil}

\date{March 12, 2026}

\begin{abstract}
Understanding thermal properties of materials is fundamental to technological applications and to discovering new phenomena. In particular, advances in experimental techniques such as cold-atom measurements allow the simulation of paradigmatic Hamiltonians with great control over model parameters, such as the Hubbard model. One aspect of this model which is not much explored is the behavior of the specific heat as a function of density. In this work, we perform Determinant Quantum Monte Carlo simulations of the Hubbard model interpolating between the square and triangular lattices to analyze the specific heat as the filling, interaction, and temperature of the system are changed. We found that, with strong correlations, the specific heat presents a three-maxima structure as a function of filling, with local minima between them. This effect can be explained by a decomposition of kinetic and potential contributions to the specific heat, demonstrating interesting phenomena away from the commonly studied half-filling regime.
Moreover, by analyzing the kinetic contribution in momentum space we show that, connected to this specific heat behavior, there is a density anomaly detected through the thermal expansion coefficient. These momentum-space quantities are accessible using cold-atom experimental measurements at multiple temperatures.
Finally, we map the location of these phenomena and connect the thermal expansion anomaly with the well-known Seebeck coefficient change of sign. Our results provide a new perspective to analyze this change of sign.
\end{abstract}

\maketitle

\section{Introduction}

Anomalous thermal properties of materials have become an intense source of research in different realms of physics. In the context of complex fluids, water presents a plethora of thermodynamic anomalies, including its unusual volume contraction from 0$^{\circ}$C to 4$^{\circ}$C at 1 atm~\cite{Kell1967}, named as density anomaly, and its high specific heat. Both, among others, are regarded as essential ingredients for life~\cite{gallo_water_2016}. Another paradigmatic example is the unexpected low thermal expansion of Fe–Ni Invar~\cite{lohaus_thermodynamic_2023}, a product of the competition between phonons and spins, which still pushes forward theoretical research and the development of materials where extreme precision is required~\cite{takenaka2012negative}. Heavy fermion compounds, such as CeRu$_2$Si$_2$~\cite{Lacerda1989}, also display a negative electronic thermal expansion as well as an unusually high Grüneisen parameter, a quantity proportional to the ratio of thermal expansion to specific heat~\cite{PhysRevB.100.054446}. Organic charge-transfer salts~\cite{manna_lattice_2010,Pustogow2022}, like the spin liquid candidate $\kappa$-(BEDT-TTF)$_2$Cu$_2$(CN)$_3$, exhibit anomalous expansivities depending on directionality, associated with a peak anomaly around 6 K which is accompanied by a hump in the specific heat. The electronic ingredient to the thermal expansion of such families of organic charge-transfer salts can be modeled with an effective Hubbard Hamiltonian~\cite{kokalj_enhancement_2015} on an anisotropic triangular lattice at half-filling, where its parameters are connected to lattice constants.

In regard to Hubbard models, the anomalous expansion was recently studied for bosons~\cite{rizzatti2020quantum} with applications to cold atom systems on which the interactions can be precisely controlled~\cite{Bloch2005,Esslinger2010,Gross2017}. It was shown theoretically that the density anomalies appear near Mott transitions and they can be traced back to residual entropies in the atomic limit~\cite{Rizzatti2019, habitzreuter_waterlike_2023}, in connection with classical Monte Carlo simulations of lattice gases~\cite{10.1063/1.4916905}. As for the specific heat, its temperature dependence has been widely analyzed in the literature of the Hubbard model at half-filling: a low and a high temperature peak exists~\cite{paiva_signatures_2001,duffy1997specific}, but the doping dependence is less explored in terms of simulations. Novel experiments measured a maximum in the specific heat of cuprates~\cite{michon_thermodynamic_2019} as a function of density. This effect has been investigated using CDMFT for the two-dimensional Hubbard model on the square lattice, predicting a maximum very close to half-filling~\cite{sordi2019specific}. Approximations on the square lattice with second neighbor hopping also show evidence of peaks~\cite{calegari_pseudogap_2013}. On the other hand, a dip in the specific heat as a function of doping has been observed in Determinant Quantum Monte Carlo (DQMC) simulations by Duffy and Moreo~\cite{duffy1997specific}.
These studies analyzed the specific heat dependence on doping for certain temperature regions and interaction strengths, but a more detailed analysis of the phenomenon is still lacking.

In this research paper, we report on simulations of the two-dimensional Hubbard model in a square lattice, presenting results for all lattice fillings and a wide range of interactions. We demonstrate that: i) as correlations between electrons increase, a three-peak structure develops in the specific heat as a function of lattice filling, at $n = 0.5$, $n = 1$, and $n = 1.5$ in the square lattice. The peak in the central density is sharp, in connection with the known behavior as a function of temperature, while the lateral peaks are broader and are altered for different lattices. ii) For strong correlations, there is a ``density anomaly'', marked by a sign change of the thermal expansion coefficient $\alpha_{\mu}$ as doping varies. We address the connection between these two effects through an analysis of the momentum distribution of the particles. iii) Finally, we clarify the mechanism associating the thermal expansion behavior with the Seebeck anomaly, a relevant transport probe recently studied in the Hubbard model~\cite{silva_effects_2023,heremans_2008,wissgott_2010,tomczak_2018}. This quantity is known to exhibit changes of sign as a function of filling $n$ and temperature $T$. Moreover, it increases significantly close to half-filling. The analysis of $\alpha_{\mu}$ sheds light on both of these characteristic behaviors.

\section{Model and Methods}

The Hubbard model~\cite{hubbard} is given by the Hamiltonian
\begin{align}
    \mathcal{H} = -&t\sum_{\braket{i, j},\sigma} \left( c_{i \sigma}^{\dagger}c_{j \sigma} + \text{h.c.} \right) - \mu \sum_{i, \sigma} n_{i \sigma} \nonumber \\
    & + U \sum_{i} \left(n_{i \uparrow} - 1/2 \right) \left(n_{i \downarrow} - 1/2\right) \;,
    \label{eq:hubbard_hamiltonian}
\end{align}
where $i$ runs over all lattice sites and $\braket{i, j}$ correspond to first neighbor sites. Operators $c^{\dagger}_{\bm{i} \sigma}$/$c_{\bm{i} \sigma}$ denote the creation/annihilation of electrons of spin $\sigma$ on a site $\bm{i}$, and the number operator is $n_{\bm{i}\sigma} \equiv c^{\dagger}_{\bm{i} \sigma} c_{\bm{i} \sigma}$. The first sum describes particle hopping processes with amplitude $t$, the second term controls the density through the chemical potential $\mu$, and the last term describes an on-site repulsive interaction $U$.

The Hamiltonian of Equation~\ref{eq:hubbard_hamiltonian} in bipartite lattices, such as the square lattice, has particle-hole symmetry. That is, $n(\mu) = 2 - n(-\mu)$. At $\mu = 0$ the system is at half-filling, $n(T, U, \mu = 0) = 1$, and its ground state is a Mott insulator. As $U$ hinders doubly occupied sites, a chemical potential $\mu \sim U$ is required to add another particle to the system, leading to the formation of a Mott plateau around half-filling~\cite{vsimkovic_2020,bonvca_2003,khatami_2011,mikelsons_2009}.

Our simulations were performed using the Determinant Quantum Monte Carlo (DQMC) method~\cite{becca2017quantum,white_1989,hirsch_1985,hirsch_1983,blankenbecler_1981,santos_introduction_2003}, an unbiased technique. It uses a discrete auxiliary field to break down the interaction term of the $d$-dimensional many-body problem into a non-interacting $(d+1)$-dimensional Hamiltonian with $L$ imaginary time slices, introducing an error of $\mathcal O(\Delta \tau ^2)$, where $\beta = L \Delta \tau$ is the inverse temperature. Our results are obtained from 10 independent simulations for each parameter set, each simulation with 2000 equilibration steps and up to 50000 measurement steps, choosing $\Delta \tau \leq 0.1$ to ensure the Trotter error is smaller than the statistical error from the Monte Carlo sampling. The lattice size is $10 \times 10$ to calculate thermodynamic quantities. We set $t = 1$ as our energy scale. The lattice constant is also set to unity.

To determine the specific heat, we numerically differentiate the internal energy at constant density using finite differences:
\begin{equation}
    c_n = c_{\mu} - \left( \frac{\partial E}{\partial \mu} \right)_{T} \frac{\alpha_{\mu}}{n^2 \kappa_T} \;,
    \label{eq:c_n}
\end{equation}
where $c_\mu = \left( \frac{\partial E}{\partial T}\right)_{\mu}$ is the specific heat at constant chemical potential, $\alpha_{\mu} = \left( \frac{\partial n}{\partial T}\right)_{\mu}$ is the thermal expansion coefficient, $\kappa_T = \frac{1}{n^2} \left(\frac{\partial n}{\partial \mu}\right)_T $ is the isothermal compressibility, $E(T, \mu)$ is the internal energy of the system, and $n \equiv \braket n$ is the average number of particles per lattice site (see Appendix~\ref{appendix:cn_derivation} for the derivation of the equation).

In the DQMC calculations, the momentum distribution of the density at a temperature $T$ can be calculated from the Fourier Transform of equal-time Green's functions
\begin{equation}
    n_{\bm k} = \braket{c^{\dagger}_{\bm{k}} c_{\bm{k}}} = 1 - G_{\bm k} \; .
\end{equation}
Using finite-differences for different temperatures, we obtain $\left( \frac{\partial n_{\bm{k}}}{\partial T} \right)_{\mu}$. To improve the resolution when plotting quantities in momentum space, this is calculated for a $24 \times 24$ lattice.

\section{Results}

\begin{figure*}
    \centering
    \includegraphics{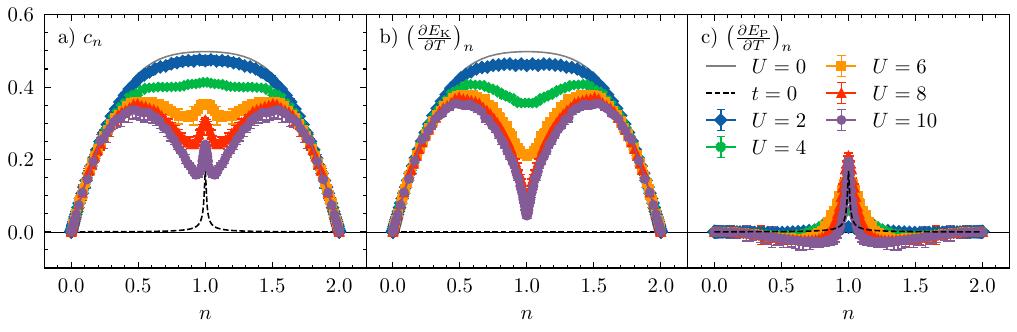}
    \caption{a) Specific heat at constant density $c_n$ as a function of lattice filling for $T = 1$ and various values of interaction $U$. b) and c) are the kinetic and potential energy contributions to the specific heat for the same temperature and interaction strengths. Results are for the $10 \times 10$ lattice.}
    \label{fig:cn_T1}
\end{figure*}

The specific heat $c_n$ as a function of doping $n$ is exhibited in Figure~\ref{fig:cn_T1} a). In the weak-coupling regime, it has a maximum at $n=1$. As the interaction $U$ becomes stronger and beyond $U \approx 6 $, a three-peak structure emerges: two lateral peaks around $n=0.5$ and $n=1.5$, and a central maximum at half-filling.
Enhancing the on-site repulsion makes the central peak sharper, and the local minima around it move towards half-filling.

In both analytically solvable scenarios, $U = 0$ (non-interacting case) and $t = 0$ (atomic limit), such a triple-peak structure is absent. Therefore, the reported behavior emerges from the electronic correlations and the interplay between $U$ and $t$. In experimental measurements, specific heat peaks at finite doping have been detected for some materials~\cite{michon_thermodynamic_2019}.

At half-filling, there is a well-known two-peak structure for $c_n(T)$~\cite{paiva_signatures_2001}: a spin peak at $T \sim J = 4t^2/U$ and a charge peak $T \sim U$, with these expressions valid only for high $U$. At low $U$, the charge peak is mostly constant at $T \sim 1$, while the spin peak exponentially decreases as $U \rightarrow 0$ and we are far from the Heisenberg regime.
Our simulations remain in temperatures above the spin scale for all $U$.
In particular, for the simulations at low $T$ and strong $U$, we have $J \lesssim T \ll U$, and double occupancy is mostly suppressed.
Hence, as $n$ is tuned away from half-filling, the main contribution to the energy should be from the hopping term, that is, the number of electron--hole bonds.

\begin{figure*}
    \centering
    \includegraphics{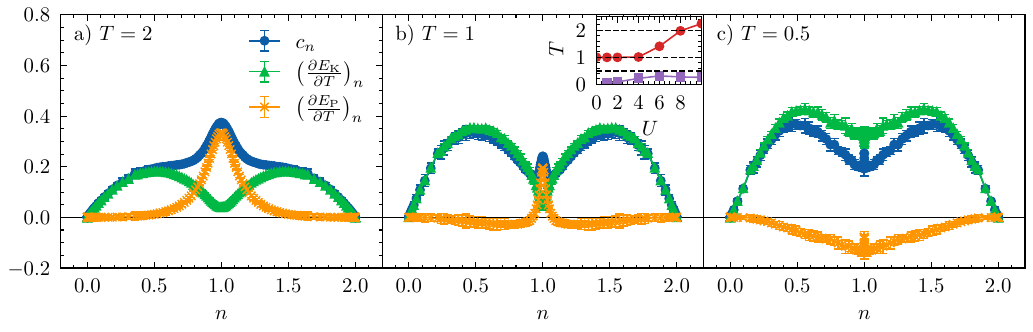}
    \caption{Specific heat and kinetic/potential contributions for $U = 10$ at temperatures a) $T = 2$, b) $T = 1$ and c) $T = 0.5$. The inset in b) is a $T \times U$ ($0 \leq U \leq 10$, $0\leq T \leq 2.5$) plot at half-filling with dashed lines showing the temperatures for which we display specific results in a), b), and c). Red circles and purple squares represent, respectively, the high and low $T$ peaks in the specific heat as a function of temperature, with data as a function of temperature of the inset extracted from reference~\cite{paiva_signatures_2001}. Results are for the $10 \times 10$ lattice.
    }
    \label{fig:cn_decomposition}
\end{figure*}

The system's average energy $E$ per lattice site can be written as the sum of the kinetic part
$E_K = \sum_{\bm{k}} \varepsilon_{\bm{k}} n_{\bm{k}} $, where $\varepsilon_{\bm{k}}$ is the lattice dispersion relation and $n_{\bm{k}}$ is the momentum distribution of particles,
and the potential term $E_P = U d$, where $d$ is the average double occupancy $d=\braket{n_\uparrow n_\downarrow}$.
Following this idea, we decompose the specific heat into its kinetic and potential energy contributions, as shown in Figure~\ref{fig:cn_T1} b) and c) respectively. The kinetic energy portion $\left(\frac{\partial E_\text{K}}{\partial T}\right)_{n}$ is nearly independent of the interaction
for low densities $0 \leq n \leq 0.5$. Thus, this regime resembles the non-interacting scenario, with electrons being too spread out for the $U$ term to be relevant.
For $0.5 < n < 1$, the repulsion effects become important (double occupancy starts to build up), and the kinetic energy contribution starts to decrease as the interaction increases. Hence, for large $U$ we have a maximum around $n = 0.5$ (and $n = 1.5$, by symmetry), whereas for weak interactions there is a single maximum at half-filling.

The potential energy contribution $\left(\frac{\partial E_\text{P}}{\partial T}\right)_{n}$ is negative in an interval $0 \leq n \lesssim n^{*}(U)$, where $n^{*}(U)$ is the location of the minimum for each interaction $U$, and presents a prominent positive peak when there is one fermion per site. Since the potential energy is proportional to the average double occupancy, its contribution to $c_n$ is governed by $\left(\frac{\partial d}{\partial T}\right)_{n}$. That is, in the low-density regime, double occupancies break up as temperature rises. It is only for $n^{*}(U) < n \leq 1$ that we observe an increase in the potential contribution, when double occupancy is favored by temperature. This is due to the large number of single-occupied sites in the lattice: thermal fluctuations will likely take an electron to a site that is already occupied, greatly enhancing the potential energy at this temperature range.
The antagonistic behavior of kinetic and potential portions close to $n = 1$ yields the local minimum of $c_n$, as shown in Figure~\ref{fig:cn_T1} a).

In Figure~\ref{fig:cn_decomposition} we plot the total $c_n$ along with its kinetic and potential contributions for $U = 10$, considering three different temperatures: a) $T = 2$, b) $T = 1$, and c) $T = 0.5$. Concerning the higher temperature $T = 2$, $c_n$ displays only a central peak, illustrated in Figure~\ref{fig:cn_decomposition} a) for this particular repulsion value. This temperature is away from the $T \sim J$ region, leading to a decreased kinetic energy contribution and no lateral maxima in the total $c_n$.
As the temperature reaches the region $J \lesssim T$, lateral peaks at $n = 0.5$ and $n = 1.5$ start to develop according to Figure~\ref{fig:cn_decomposition} b). Their magnitude is slightly enhanced at even lower temperatures as shown in Figure~\ref{fig:cn_decomposition} c), even though we are still at $J \lesssim T$.

At the lowest temperature we considered, the potential energy contribution to $c_n$ may be negative. This implies that $\left(\frac{\partial d}{\partial T}\right)_{U, n}<0$, an effect particularly important in cold atoms experiments exploring adiabatic cooling: since $ \left(\partial s/ \partial U \right)_{T, n} =- \left( \partial d / \partial T \right)_{U, n}$, then $c_n \left( \partial T / \partial U \right)_{s, n} = T \left( \partial d / \partial T \right)_{U, n}$ [37] and an increase in $U$ can be used to decrease the temperature at constant entropy if $\left( \partial d / \partial T \right)_{U, n} < 0$.
At half-filling, this has been studied for square \cite{paiva_2010,khatami_2011}, kagome \cite{andressa_2023}, and honeycomb \cite{tang_2012,tang_2013} lattices. The results of panels b) and c) show that this can happen for the whole doping range.

This counterintuitive behavior is claimed to be analogous to the
Pomeranchuk effect in liquid ${}^3\mathrm{He}$~\cite{werner_interaction-induced_2005}. For the Hubbard model, since the (spin) entropy is larger in a localized state than when the fermions form a Fermi liquid, it is favorable to increase the degree of localization upon heating. At low $T$, the system behaves as a coherent Fermi liquid with small entropy,
while at higher temperatures, local moments increase with $T$, favoring electron localization, creating a Pomeranchuk-like tendency toward the higher-entropy local-moment
regime.

The inset in Figure~\ref{fig:cn_decomposition} b) is a $T \times U$ plot at half-filling with dashed lines showing the temperatures for which we display results. Red circles and purple squares represent, respectively, the high and low $T$ peaks in the specific heat as a function of temperature. Hence, at our higher temperature, $T = 2$, we cross the high $T$ peak for large $U$. At our intermediate temperature $T = 1$, we cross the high $T$ peak for small $U$. Our lower temperature, $T = 0.5$, does not cross any peak as a function of temperature at half-filling, hence guaranteeing we are in the $J \lesssim T$ region for $T = 0.5$ and $T = 1$, while $T = 2$ is above this spin regime for most interactions.

Therefore, the decomposition of $c_n$ for $J \lesssim T$ highlights the different nature of the central and lateral peaks. The lateral, broader ones come from the strong decrease in the kinetic energy due to reduced carrier mobility.
The central sharp peak is due to an increase in double occupancy for small to intermediate temperatures, and decreases as $T$ decreases due to the formation of local moments. At $T \sim U$ only the central peak is detected, and at very low temperature, we expect all peaks to vanish.

\begin{figure}
    \centering
    \includegraphics{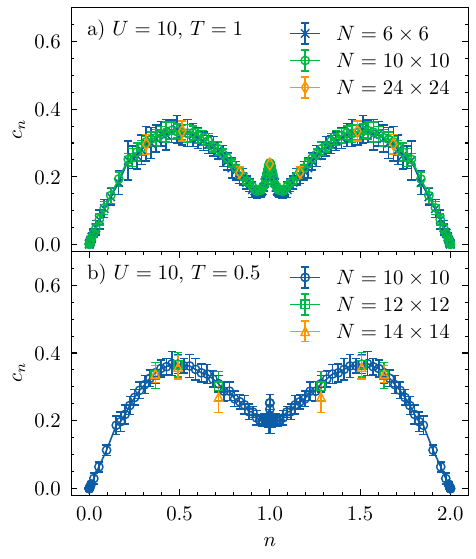}
    \caption{Specific heat as a function of lattice filling $n$ for $U = 10$ and different lattice sizes at a) $T = 1.0$ and b) $T = 0.5$.}
    \label{fig:cn_U10_changeSize}
\end{figure}

We analyze finite-size effects of the specific heat in Figure~\ref{fig:cn_U10_changeSize}. We fixed $U = 10$ and $T = 1$ in panel a), and considered lattice sizes from $6 \times 6$ to $24 \times 24$. Results coincide within error bars, indicating that our analysis in $10 \times 10$ lattices provides a good estimate of the thermodynamic limit results. For the lower temperature of $T = 0.5$ shown in panel b) of Figure~\ref{fig:cn_U10_changeSize}, the specific heat is also in good agreement within error bars. We note that even the half-filling peak remains the same for large sizes, although it could be expected to increase in the thermodynamic limit for lower temperatures than considered here, due to the known results at $n=1$~\cite{paiva_signatures_2001}. In our results, this does not happen because the temperature $T = 0.5$ is situated between the peaks as a function of temperature, as shown in the inset of Figure~\ref{fig:cn_decomposition} b).

\begin{figure*}
    \centering
    \includegraphics{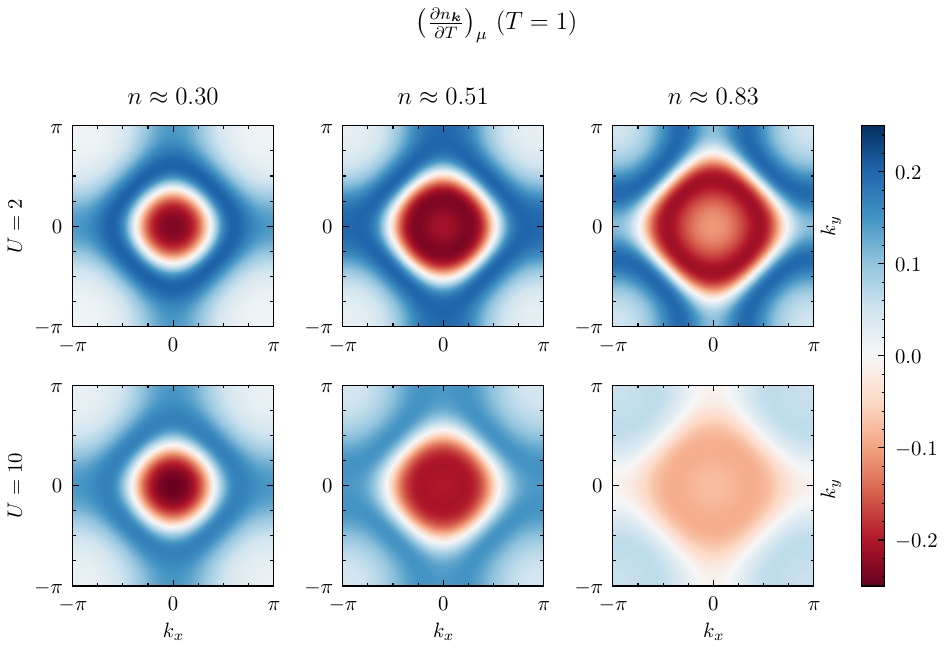}
    \caption{The momentum distribution of $\left( \frac{\partial n_{\bm{k}}}{\partial T} \right)_{\mu}$ in the first Brillouin zone for interactions $U = 2$ (upper panels) and $U = 10$ (lower panels), and fillings $n \approx 0.3$ (left panels), $n \approx 0.51$ (central panels) and $n \approx 0.83$ (right panels). The temperature is $T = 1$ and the lattice size is $24 \times 24$, with a cubic spline interpolation between points.}
    \label{fig:dnkdT}
\end{figure*}

\begin{figure*}
    \centering
    \includegraphics{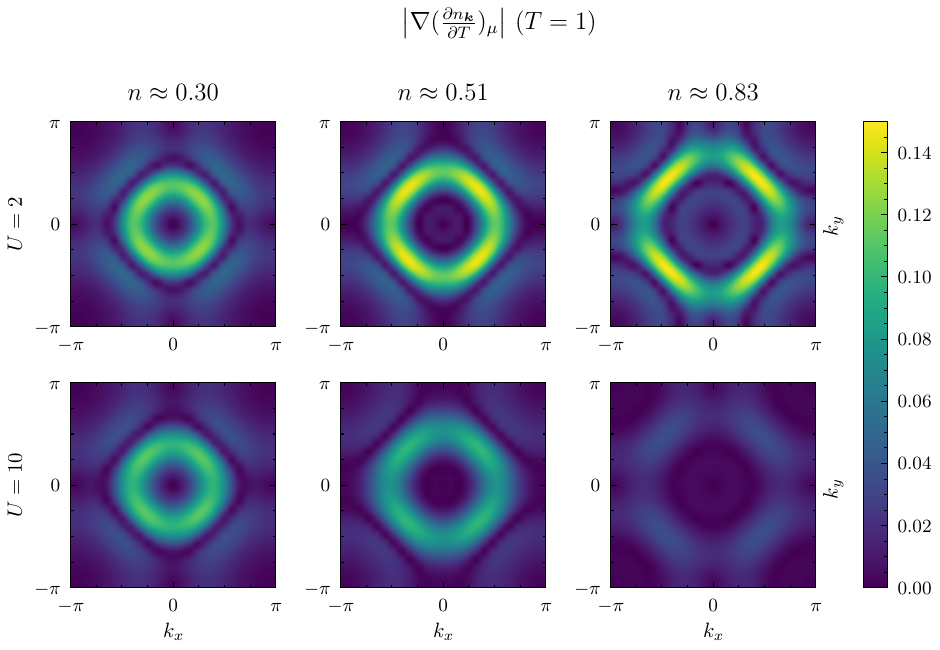}
    \caption{The gradient of the momentum distribution of $\left( \frac{\partial n_{\bm{k}}}{\partial T} \right)_{\mu}$ in the first Brillouin zone for interactions $U = 2$ (upper panels) and $U = 10$ (lower panels), and fillings $n \approx 0.3$ (left panels), $n \approx 0.51$ (central panels) and $n \approx 0.83$ (right panels). The temperature is $T = 1$ and the lattice size is $24 \times 24$, with a cubic spline interpolation between points.}
    \label{fig:grad_dnk_dT_fixmu}
\end{figure*}

Inasmuch as the decrease of $c_n$ in the region $0.5 < n < 1$ is dominated by the kinetic contribution and
\begin{equation}
    \left(\frac{\partial E_\text{K}}{\partial T}\right)_{n} = \sum_{\bm{k}} \varepsilon_{\bm{k}} \left(\frac{\partial n_{\bm{k}}}{\partial T}\right)_{n} \; ,
\label{eq:cn_kin_contribution_k_decomposition}
\end{equation}
we analyze in Figure~\ref{fig:dnkdT} the temperature derivative of $n_{\bm k}$ within the first Brillouin zone. The six panels cover weak and strong couplings as the density varies. At a fixed filling, large $U$ increases correlations in the real domain and delocalizes electrons in momentum space. Consequently, correlations induce a damping in the magnitude of $\left( \frac{\partial n_{\bm{k}}}{\partial T}\right)_{n}$. This effect gets more pronounced as one approaches the Mott state, where the system gets fully localized in real space. This is more clearly seen in the absolute value of the gradient, shown in Figure~\ref{fig:grad_dnk_dT_fixmu}.

The lower sensitivity to temperature at high density and strong coupling is reflected in the kinetic energy derivative with temperature of Equation~\ref{eq:cn_kin_contribution_k_decomposition}, which is lower for $n = 0.83$ than for $n = 0.51$. This explains the reduction of $c_n$ beyond the quarter filling. Commonly addressed in cold atoms experiments, the momentum distribution~\cite{Maka2011} imaging is accessible through time-of-flight measurements~\cite{DeMarco1999,Greiner2003, holten2022observation}, but measurements at multiple temperatures~\cite{cocchi_2017_entropy_measurement} would have to be performed to calculate the temperature derivative $\partial n_{\bm k} / \partial T$.
Double occupancy measurements for fermionic atoms with atomic resolution are available through quantum gas microscopy~\cite{Cheuk2015,Parsons2015,Edge2015,Haller2015}, again requiring measurements at different temperatures for a numerical differentiation.

\begin{figure}
    \centering
    \includegraphics[]{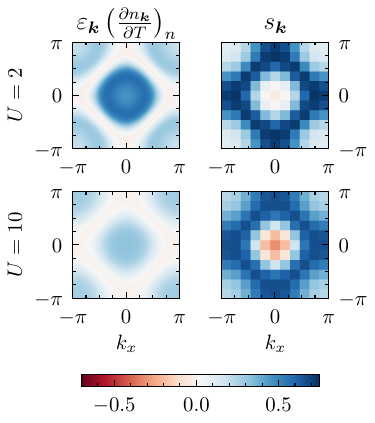}
    \caption{Left panels show the momentum space contributions to $\varepsilon(\bm k) \left( \frac{\partial n_{\bm{k}}}{\partial T} \right)_{n}$. Right panels shows the momentum space contributions to the entropy. The interaction is $U = 2$ (upper panels) and $U = 10$ (lower panels), all of them for a filling $n \approx 0.83$. The temperature is $T = 1$ and the lattice size is $24 \times 24$ on the left panels, with a cubic spline interpolation between points. The right panels are calculated for a $10 \times 10$ lattice due to $\mu$ integration needed to find $s$.}
    \label{fig:ek_dnk_dT_fix_n0.83n}
\end{figure}

While Figures~\ref{fig:dnkdT} and~\ref{fig:grad_dnk_dT_fixmu} highlight the reduced temperature sensibility for strong interactions at $n \approx 0.8$, we need $\varepsilon(\bm k) \left( \frac{\partial n_{\bm{k}}}{\partial T} \right)_{n}$ to understand which regions contribute to the specific heat (see Appendix~\ref{appendix:comparison} for a numerical comparison of $\left( \partial n_{\bm k} / \partial T \right)_{\mu}$ and $\left( \partial n_{\bm k} / \partial T \right)_{n}$, they are qualitatively the same). We plot this in the left panels of Fig.~\ref{fig:ek_dnk_dT_fix_n0.83n}, for $n \approx 0.83$ at $T = 1$. For $U = 2$, the largest contribution comes from small $k$s. Increasing the interaction to $U = 10$ weakens this central feature, while the border regions remain mostly unaltered. Hence, the lower specific heat is due to a decrease in the thermal sensitivity of low $k$ electrons at the center of the Brillouin zone.

Note that the entropy is related to $\left(\frac{\partial n}{\partial T}\right)_{\mu}$:
\begin{align}
    s(T, \mu) = \int_{-\infty}^{\mu} \left( \frac{\partial n}{\partial T} \right)_{\mu'} d\mu'
\end{align}
Since $n = \sum_{\bm k} n_{\bm k}$, we can decompose it as
\begin{align}
    s(T, \mu)
    &= \sum_{\bm k} \int_{-\infty}^{\mu} \left( \frac{\partial n_{\bm k}}{\partial T} \right)_{\mu'} d\mu' \\
    &= \sum_{\bm k} s_{\bm k} (T, \mu)
\end{align}
In the right panels of Fig.~\ref{fig:ek_dnk_dT_fix_n0.83n} we show the momentum space decomposition of the entropy, $s_{\bm k}$ at a density $n \approx 0.83$ and temperature $T = 1$ for the weakly interacting ($U = 2$) and strong coupling ($U = 10$) regimes. Due to the $\mu'$ integration required, we are only able to perform this calculation for $10 \times 10$ systems. Surprisingly, at strong coupling, we see a negative contribution to the entropy from the electrons at the center of the Brillouin zone, while the positive entropy contributions come from the Fermi surface region. This is consistent with the decreasing entropy as a function of $n$ observed in other works~\cite{bonvca_2003,silva_effects_2023} and shown for various temperatures in Appendix~\ref{appendix:entropy}. The decomposition results of Fig.~\ref{fig:ek_dnk_dT_fix_n0.83n} do not imply a negative thermodynamic entropy: $s(T, \mu) = \sum_{\bm k} s_{\bm k} > 0$ must hold as expected.

\begin{figure}
    \centering
    \includegraphics{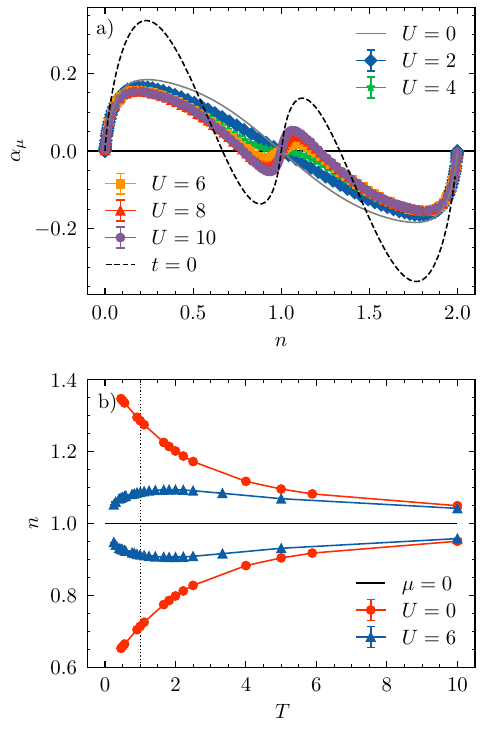}
    \caption{a) Thermal expansion $\alpha_{\mu}$ as a function of density for $T = 1$, considering interactions ranging from the weak to the strong coupling limit. b) Density as a function of temperature, showing a case with a regular density behavior (red circles, $U = 0$) and an anomalous behavior (blue triangles, $U=6$). The continuous black line is the half-filling density. The vertical dotted line is $T = 1$, corresponding to the fixed temperature in panel a). Results are for the $10 \times 10$ lattice.}
    \label{fig:alpha_mu_T1}
\end{figure}

The thermodynamic quantity associated with $\left( \partial n_{\bm k} / \partial T \right)_{\mu}$ is its momentum space sum: the thermal expansion coefficient
\begin{equation}
    \alpha_{\mu} = \sum_{\bm{k}} \left( \frac{\partial n_{\bm{k}}}{\partial T} \right)_{\mu}
\end{equation}
at constant $\mu$. Figure~\ref{fig:alpha_mu_T1} a) shows, for different lattice fillings and interaction strengths at a temperature $T = 1$, a trivial crossing at $\alpha_{\mu} = 0$ and $n = 1$ due to particle-hole symmetry of the square lattice, since $n=1$ at $\mu=0$ for any temperature and interaction. For weak interactions (up to $U \approx 4$), we observe $\alpha_{\mu}>0$ below half-filling and $\alpha_{\mu}<0$ above it. This describes the regular response of the thermal expansion predicted by the non-interacting solution. Indeed, if $n<1$, raising the temperature increases the average occupation number at fixed chemical potential, whereas increasing the temperature tends to decrease the density when $n>1$. As $U$ becomes larger (beyond $U \approx 6$), the thermal expansion takes on negative values just below the half-filling and positive ones just above it. This composes the signature of the anomaly: $\alpha_{\mu} < 0$ for $n < 1$ and $\alpha_{\mu} > 0$ for $n > 1$. The reported anomalous sign change persists up to the atomic limit, which can be solved analytically.

Figure~\ref{fig:alpha_mu_T1} b) shows the density as a function of temperature at $|\mu| = 1$. The red curves portray the non-interacting case $U = 0$, where the particle's thermal response is regular. The blue curves correspond to the intermediate coupling scenario $U = 6$ for which the anomaly is observed at low temperatures $T\lesssim 2$. This kind of effect has been widely studied for classical liquid systems with short-range interactions at competing distance scales, most notably water, and on Monte Carlo simulations of classical lattice models~\cite{de2005density,franzese2002theory, barbosa2011thermodynamic}. In the context of Hubbard physics, the anomalous thermal response arises from the strongly correlated particles' motion near the Mott states \cite{vsimkovic_2020,bonvca_2003,khatami_2011,mikelsons_2009}. Recently, some of us proposed such an effect for a system of bosons~\cite{Rizzatti2019,rizzatti2020quantum}. Thus, these DQMC results for the Hubbard model demonstrate that it can also happen for Fermi statistics.

The thermal expansion also influences thermoelectric effects, such as the generation of a particle current by a temperature gradient. Indeed, the last term of Equation~\ref{eq:c_n} corresponds to the Seebeck coefficient according to the Kelvin formula~\cite{shastry_2008,peterson_2010},
\begin{equation}
    \frac{\alpha_{\mu}}{n^2 \kappa_T} = \left(\frac{\partial s}{\partial n}\right)_{T} = q \; S_{\text{Kelvin}} \; .
    \label{eq:Kelvin}
\end{equation}
where $q$ is the carrier charge.
In particular, $S_{\text{Kelvin}}$ arises from the slow limit of the general Kubo formula for thermopower, that is, $\lim \{q_x \rightarrow0, \omega \rightarrow0\} \; S(\bm q, \omega)$.
We set $q = -1$ as the unit of charge in the rest of the text.

In the context of the Hubbard model, it has been established that the Kelvin approximation is one of the best available~\cite{shastry_2008}, even being in excellent quantitative agreement with experiments for some materials~\cite{wang_quantitative_2023}.
This is due to the fact that, at the high temperatures and strong interactions, transport is incoherent. At low temperatures not accessible to our simulations the thermopower is better described by the high-frequency limit since there is coherent transport~\cite{tremblay_thermopower_2013}.

Because the denominator of Equation~\ref{eq:Kelvin} is always positive due to thermodynamic stability in the single-band Hubbard model~\cite{hubbardmodel_stability}, a sign change of the Seebeck coefficient must be induced by a sign change of $\alpha_{\mu}$.
Namely, the reported Seebeck anomaly~\cite{silva_effects_2023} regarding its sign, and hence carrier charge, is controlled by the thermal expansion, with its magnitude modulated by the isothermal compressibility.
This connection can be conceptually visualized as follows: a given temperature gradient $\delta T > 0$ will induce a density change $\delta n > 0$, as long as $\alpha_{\mu}$ is positive. This charge variation will induce an electric field in the same direction, and hence, we get a negative coefficient. On the other hand, an opposite response of density $\delta n < 0$, due to a negative $\alpha_{\mu}$, will establish an electric field in the opposite direction, yielding a positive Seebeck response.
Near half-filling and for strong interactions, $n^2 \kappa_T$ goes to zero~\cite{silva_effects_2023}, greatly enhancing the Seebeck coefficient. But the thermal expansion coefficient is zero at $n = 1$ due to particle--hole symmetry, leading to a zero Seebeck
Microscopically, the change in carrier number above (below) half-filling is connected to the restructuring of the Fermi surface, which shifts from hole-like  (electron-like) to electron-like (hole-like) above (below) half-filling~\cite{osborne_broken_2021}.

Studies of classical lattice models connected a change of sign in $\alpha_{\mu}$ to residual entropies~\cite{10.1063/1.4916905}, as well as the results for bosons~\cite{Rizzatti2019,rizzatti2020quantum}. Although the ground state of the Hubbard model has no residual entropy in the thermodynamic limit, at finite temperature, the entropy as a function of lattice filling $s(n)$ has maxima away from half-filling for strong correlations~\cite{silva_effects_2023}.  From Equation~\ref{eq:Kelvin}, we can see that $\alpha_{\mu} = 0$ implies $(\partial s / \partial n)_T = 0$. That is, the change of sign of $\alpha_{\mu}$ happens right at the entropy maxima. Conversely, the same is true for the Seebeck coefficient: its anomalous change of sign happens at the maxima in the entropy.

\begin{figure}
    \centering
    \includegraphics{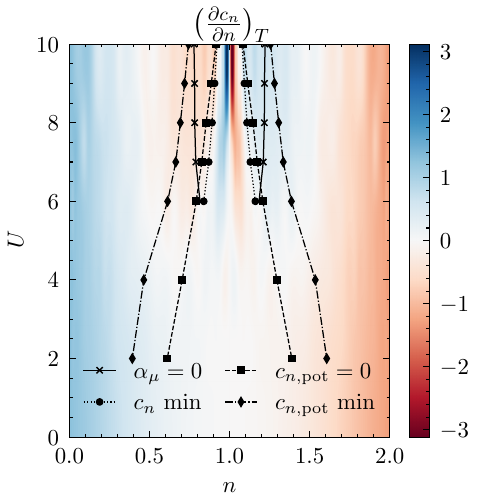}
    \caption{Heat map of $\left(\frac{\partial c_n}{\partial n}\right)_T$ for  $T = 1$ in the $n$ versus $U$ plane. Simulations were performed for  $U \in \{2, 4, 6, 7, 8, 9, 10\}$ to produce an interpolation. Crosses mark states with $\alpha_{\mu} = 0$. Filled circles denote where $c_n$ is minimum, filled squares represent $c_{n, \text{pot}} = 0$, and black diamonds indicate the loci of minimum $c_{n, \text{pot}}$. Results are for the $10 \times 10$ lattice.}
    \label{fig:heatmap_T1}
\end{figure}

To tie up all of these results, Figure~\ref{fig:heatmap_T1} presents a heat map of $\left(\partial c_n / \partial n\right)_T$, marking the locations of the specific heat and density anomalies in the $n$ versus $U$ plane.
Black diamonds indicate where the potential energy contribution of $c_n$, $c_{n, \text{pot}}$, is minimum, in connection with adiabatic cooling. As the repulsion increases, this minimum shifts towards densities closer to the half-filling.
Black squares denote where $c_{n, \text{pot}}$ changes sign, pointing out the density values where double occupancy begins to be favored. As the interaction is increased, this curve crosses the $\alpha_{\mu} = 0$ line at intermediate values of $U$, moving towards the half-filling.
Filled black circles and crosses show, respectively, the loci of minimum values of $c_n$ and $\alpha_{\mu}=0$ (or  $\left(\frac{\partial s}{\partial n}\right)_{T} = 0$). For $U \approx 6$ these crosses and circles are very close, as well as the change of sign in the potential energy contribution.

Increasing the repulsion $U$ shifts the $c_n$ minimum to densities $n \approx 1$, while the zero crossing of $\alpha_{\mu}$ is shifted to densities away from half-filling.
In general, there is an intricate relation between  $c_{\mu}$, $E(\mu)$, and $\alpha_{\mu}$ with minima and maxima of each term of Eq.~\ref{eq:c_n} being interconnected in complex ways. Our data of $(\partial n_{\bm k}/\partial T)_{\mu}$ shown in Fig.~\ref{fig:dnkdT} suggests that, when this quantity is weaker over the Brillouin zone, the sum of $(\partial n_{\bm k} / \partial T)_{n}$ weighted by $\varepsilon(\bm k)$ yields a smaller $c_n$ for this system. Hence the lower value of $c_n$ around $n \approx 0.83$ in comparison with $n \approx 0.5$ at strong interactions. This qualitatively explains why the $\alpha_{\mu} = 0$ curve is contained in the $\left( \frac{\partial c_n}{\partial n} \right)_T < 0$ ($> 0$) below (above) half-filling in our results of Fig.~\ref{fig:heatmap_T1}.
We show the entropy $s(n)$ and $\alpha_{\mu} = 0$ data for other temperatures in Appendix~\ref{appendix:entropy}.

The kinetic part of the specific heat suggests carrier mobility influences the lateral maxima. Thus, we also analyzed the specific heat as a function of density when an additional hopping $t'$ is added in the $\hat x + \hat y$ direction, breaking particle-hole symmetry. For $t' = 1$, the system is equivalent to a triangular lattice.

\begin{figure*}[t]
    \centering
    \includegraphics{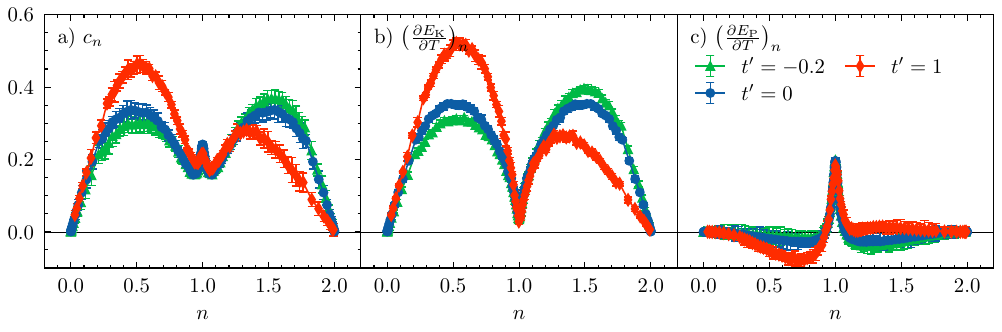}
    \caption{Specific heat decomposition for $U = 10$, $T = 1$ with an additional hopping $t'$ in the $\hat x + \hat y$ direction in the $10 \times 10$ lattice.}
    \label{fig:cn_tp_decomposition}
\end{figure*}

In Figure~\ref{fig:cn_tp_decomposition}a), we show the cases $t' = -0.2$, $t' = 0$, and $t' = 1$. At half-filling, all of them show a peak, although the intensity is reduced at $t' = 1$ compared with the $t' = 0$ case. For larger doping, the loss of particle-hole symmetry is seen: the specific heat peak at $n \approx 1.5$ is larger for the $t' = -0.2$ than the $t' = 0$ case, while the peak at $n \approx 0.5$ has the opposite behavior. For the $t' = 1$ case, the result is more intense, with the $n = 1.5$ peak shifting to lower densities and greatly diminishing. Moreover, the peak at $n = 0.5$ is significantly enhanced. For $t' = -0.2$, we observe the inverted behavior, but this specific value of hopping is too small for us to detect the lateral shift.

In panels b) and c) of Figure~\ref{fig:cn_tp_decomposition}, we display the specific heat decomposition into kinetic and potential contributions with the new hopping term. Both the shift to lower densities and the intensity decrease in the case of $t' = 1$ are driven mostly by the kinetic contribution. In the potential term, we see a stronger $(\partial d / \partial T)_n < 0$ below half-filling, but there is no adiabatic cooling above half-filling.

\section{Conclusions}

Using Quantum Monte Carlo simulations, we studied the triple-peak structure of the specific heat as a function of the filling. This effect emerges in the temperature regime $J \lesssim T \ll U$, where $J = 4t^2/U$. As the filling of the lattice is increased below half-filling, a decreasing kinetic energy leads to a lower specific heat at constant density with increasing $U$, meaning that it is easier to heat up the system for larger interaction strengths. The maximum threshold of $n = 0.5$ in the specific heat does not depend much on the interaction $U$ for the square lattice. However, this was considerably altered in the triangular lattice, with differing heights and positions of maxima.

The reported effect complements the half-filling $c_n(T)$ known peak structure. We explained it by analyzing the momentum space distribution $n_{\bm k}$ of the interacting fermions, showing the low--$k$ electrons become less sensitive to temperature at the interactions and lattice fillings where $c_n(n, U)$ behaves differently from the non-interacting case.
The distribution $n_{\bm k}$ is experimentally measurable in cold atom experiments.
Moreover, we detected an enhanced negative potential energy contribution at lower temperatures. Due to double occupancy decreasing with temperature, this leads to adiabatic cooling at a wide range of densities.

Our data suggests a connection between the specific heat $c_n(n)$ decreasing (increasing) below (above) half-filling to $(\partial n_{\bm k} / \partial T)$.
This quantity is related to a density anomaly, widely explored in the context of complex fluids and materials engineering.
Its signature in the Hubbard model is $\alpha_{\mu} < 0$ below half-filling and $\alpha_{\mu} > 0$ above it, which are found for strong correlations $U \gtrsim 6$.

The density response also regulates thermoelectric properties: the sign of the Seebeck coefficient (in the long wavelength and static limits) is determined by the thermal expansion coefficient, as well as maxima in the entropy as a function of density.
Therefore, Seebeck and density anomalies are tied together in the regime where the Kelvin formula for the thermopower is valid.
Further research in this direction could analyze additional interaction scales, such as the extended Hubbard model, in analogy to classical systems with multiple potential wells or steps~\cite{rizzatti_core-softened_2018}.

Understanding thermal anomalies is a crucial step to stretching the limits of cooling processes. For instance, the Mpemba effect, which is still the subject of debate in the case of water, is being studied with quantum simulators~\cite{PhysRevLett.133.010402}: systems which are completely different are connected by how fast cooling or heating can happen.
Furthermore, we note that a finite-temperature approximant for the Von Neumann entropy has been reported to show maxima as a function of density~\cite{silva2025single}. A possible avenue of future research is to understand how these maxima relate to the ones shown in our work.

\begin{acknowledgments}
We thank Natanael C. Costa and Rodrigo G. Pereira for discussions about this work.
We thank CENAPAD-SP (Centro Nacional de Processamento de Alto Desempenho em São Paulo), projeto UNICAMP/FINEP-MCTI for computational time for the DQMC calculations.
M.A.H. thanks CAPES for funding and IIP-UFRN, where part of this work was completed (Grant No. 1699/24 IIF-FINEP).
M.C.B. thanks CNPq for grants 403427/2021-5 and 311707/2021-1.
Financial support from Fundação Carlos Chagas Filho de Amparo à Pesquisa do Estado do Rio de Janeiro, grant number E-26/204.308/2021 (W.C.F.S.),
E-26/200.959/2022 (T.P.), E-26/210.100/2023 (T.P.), and E-26/210.781/2025 (T.P.), T. P.  also acknowledges support from CNPq grant numbers 308335/2019-8, 403130/2021-2,  and 442072/2023-6.
\end{acknowledgments}

\bibliography{apssamp}

\newpage
\appendix
\renewcommand{\thefigure}{\thesection\arabic{figure}}
\setcounter{figure}{0}

\section{$c_n$ derivation}
\label{appendix:cn_derivation}

The specific heat at constant density is defined as
\begin{equation}
    c_n = \left( \frac{\partial E}{\partial T} \right)_{n} \; .
\end{equation}
Since we are working in the Grand Canonical Ensemble, $E = E(T, n) = E(T, n(T, \mu))$, hence
\begin{equation}
    \left( \frac{\partial E}{\partial T} \right)_{n} = \left( \frac{\partial E}{\partial T} \right)_{\mu} + \left( \frac{\partial E}{\partial \mu} \right)_{T} \left( \frac{\partial \mu}{\partial T} \right)_{n} \;.
\end{equation}
We use that
\begin{equation}
    \left( \frac{\partial \mu}{\partial T} \right)_{n} = -\frac{\left( \frac{\partial n}{\partial T} \right)_{\mu}}{\left( \frac{\partial n}{\partial \mu} \right)_{T}} \; .
\end{equation}
Hence,
\begin{equation}
    \left( \frac{\partial E}{\partial T} \right)_{n} = \left( \frac{\partial E}{\partial T} \right)_{\mu} - \frac{\left( \frac{\partial E}{\partial \mu} \right)_{T} \left( \frac{\partial n}{\partial T} \right)_{\mu}}{\left( \frac{\partial n}{\partial \mu} \right)_{T}} \; .
\end{equation}
Identifying
\begin{align}
    c_\mu &=  \left( \frac{\partial E}{\partial T} \right)_{\mu} \;,\\
    \alpha_\mu &= \left( \frac{\partial n}{\partial T} \right)_{\mu}  \;, \\
    \kappa_T &= \frac{1}{n^2} \left( \frac{\partial n}{\partial \mu} \right)_{T} \;,
\end{align}
we arrive at Equation~\ref{eq:c_n}.

\section{Sign problem \& error propagation}

\begin{figure}[t]
    \centering
    \includegraphics{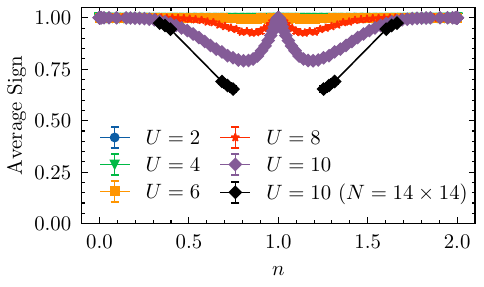}
    \caption{Average sign at $T = 0.5$ for various interactions $U$. Points are for the $N = 10 \times 10$ lattice unless the label specifies otherwise. The black points correspond to the calculated fillings for the largest lattice shown in Figure~\ref{fig:cn_U10_changeSize}. Continuous lines are a guide to the eyes.}
    \label{fig:avg_sign}
\end{figure}

The DQMC technique is unbiased but suffers from the well-known sign problem~\cite{hirsch_1985,white_1989}. This requires a reweighting of the observables:
\begin{equation}
    \braket{A} = \frac{\braket{\text{sgn} \times A}}{\braket{\text{sgn}}} .
\end{equation}
That is, instead of measuring the observable $A$, we measure $\text{sgn} \times A$ and normalize by the average sign. If the average sign $\braket{\text{sgn}}$ is close to zero, good statistical averages for the observable require exponentially longer simulations. The repulsive Hubbard model is sign-problem-free at half-filling ($\mu = 0$), but doping changes this scenario. In general, the sign problem becomes worse at lower temperatures, in larger systems, and with stronger interactions. In our work, we restricted ourselves to temperatures at which the sign problem was not severe. We show our lowest temperature $T = 0.5$ in Figure~\ref{fig:avg_sign} for the $10 \times 10$ lattice.

The DQMC simulations measure $\braket{A}$ multiple times, leading to an average $\overline{A}$ and standard deviation $\delta A$.
We can then calculate numerical derivatives through finite differences, taking into account the error bars. For instance, the temperature derivative of the internal energy and its error are calculated with
\begin{align*}
    \left(\frac{\partial E}{\partial T}\right)_{\mu} &= -\beta^2 \frac{E(\beta + 2\Delta \tau, \mu) - E(\beta - 2\Delta \tau, \mu)}{4 \Delta \tau} \\
    \delta \left(\frac{\partial E}{\partial T}\right)_{\mu} &= \beta^2 \frac{\sqrt{\delta E(\beta + 2\Delta \tau, \mu)^2 + \delta E(\beta - 2\Delta \tau, \mu)^2}}{4 \Delta \tau}
\end{align*}
Note that in DQMC there is an imaginary time dimension which is discretized in $\Delta \tau$ steps. Hence the finite difference must be applied in temperatures which are an integer multiple of $\Delta \tau$. In particular, for $T = 0.5$ we used $\Delta \tau = 0.05$ and $L = 36, 40, 44$.

\section{Entropy for various temperatures}
\label{appendix:entropy}

\begin{figure*}
    \centering
    \includegraphics{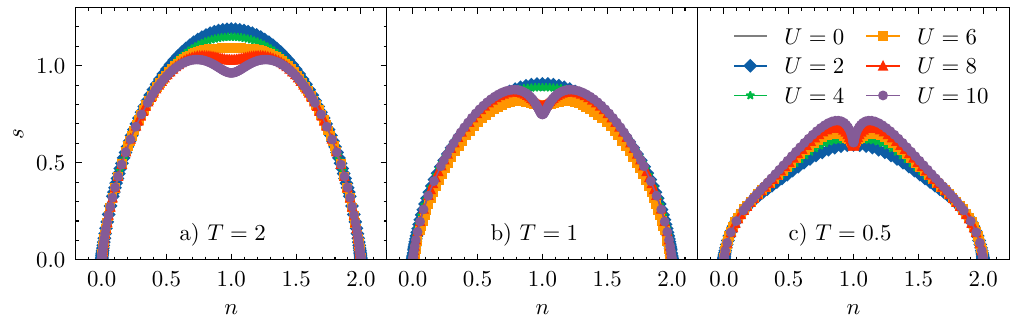}
    \caption{Entropy $s$ as a function of density $n$ for a) $T = 2$, b) $T = 1$ and c) $T = 0.5$ considering interactions ranging from the weak to the strong coupling limit. The $U = 0$ line is hidden behind the $U = 2$ case.}
    \label{fig:entropy}
\end{figure*}

Figure~\ref{fig:entropy} shows that we detect the known maxima away from half-filling for all simulated temperatures. While for $T = 2$ only $U = 8$ and $U = 10$ show maxima away from half-filling, at $T = 0.5$ this effect begins already at $U = 4$. At the extreme values of $n = 0$ and $n = 2$ the system is a band insulator, leading to zero entropy. Note that at the low temperature the $U = 10$ case has the highest entropy and $U = 0$ the lowest, while the opposite happens at $T = 2$, for all fillings. The $T = 0.5$ results were shown previously in~\cite{silva_effects_2023}.

\begin{figure}
    \centering
    \includegraphics{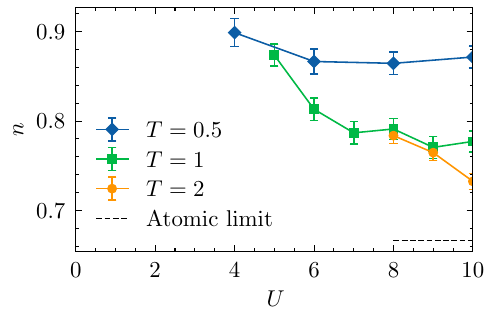}
    \caption{Location of $\alpha_{\mu} = S_{\text{Kelvin}} = (\partial s/\partial n)_T = 0$ for all the simulated temperatures and interactions, estimated by the average between densities with opposite $\alpha_{\mu}$ signs. Here we plot only $n < 1$. Error bars are estimated as the difference between densities with opposite $\alpha_{\mu}$ signs. }
    \label{fig:n_vs_U}
\end{figure}

To map the location of the peaks as temperature changes, we mark on Figure~\ref{fig:n_vs_U} the points where $\alpha_{\mu} = S_{\text{Kelvin}} = (\partial s/\partial n)_{T} = 0$, for all simulated temperatures and interactions. For the higher temperature, strong interactions ($U > 8$) are necessary for the change of sign to occur. As the temperature is lowered, the effect starts to occur at a lower value of $U$. Our results are in quantitative agreement with behavior of the peaks available in the literature, even though the previous calculation was for a smaller $4 \times 4$ system~\cite{bonvca_2003} which resulted in additional entropy peaks at other fillings.

\section{Comparison of specific heat derivative at constant density and chemical potential}
\label{appendix:comparison}

\begin{figure*}
    \centering
    \includegraphics{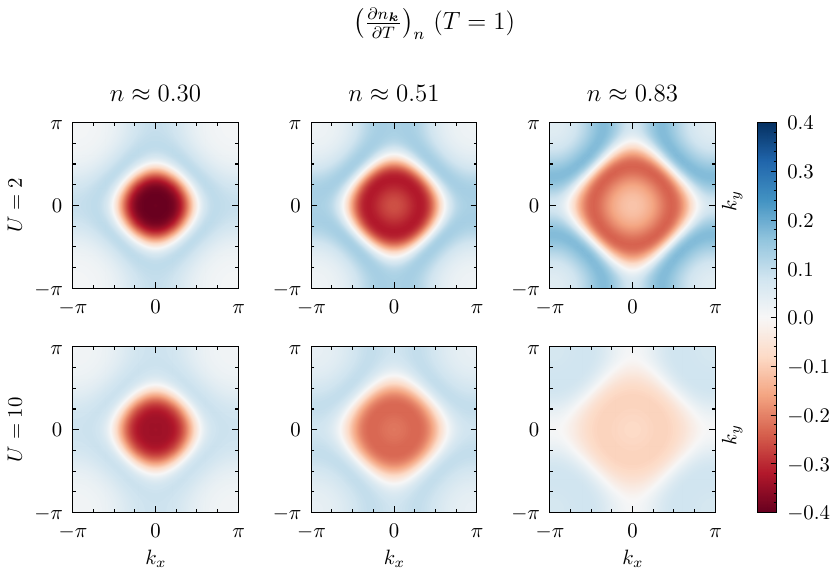}
    \caption{The momentum distribution of $\left( \frac{\partial n_{\bm{k}}}{\partial T} \right)_{n}$ in the first Brillouin zone for interactions $U = 2$ (upper panels) and $U = 10$ (lower panels), and fillings $n \approx 0.3$ (left panels), $n \approx 0.51$ (central panels) and $n \approx 0.83$ (right panels). The temperature is $T = 1$ and the lattice size is $24 \times 24$, with a cubic spline interpolation between points.}
    \label{fig:dnkdT_fixn}
\end{figure*}

We show in Figure~\ref{fig:dnkdT_fixn} the momentum-space temperature derivative of the occupation $n_{\bm k}$ at constant density $n$. In a similar manner as shown for the constant-$\mu$ derivative, the strong interaction at $n \approx 0.83$ leads to a low temperature sensitivity in comparison with the other panels.

\end{document}